\documentclass[11pt,letterpaper,reqno]{amsart}

\author{Tomoki Ohsawa and Anthony M.~Bloch}
\address{Department of Mathematics, University of Michigan, 530 Church Street, Ann Arbor, Michigan 48109-1043}
\email{ohsawa@umich.edu, abloch@umich.edu}
\title{Nonholonomic Hamilton--Jacobi Equation and Integrability}

\usepackage{amsmath,amsthm,amssymb,graphicx}
\usepackage[margin=1in]{geometry}

\usepackage{bigstrut,multirow}

\usepackage[numbers,sort&compress]{natbib}

\usepackage[colorlinks=false]{hyperref}

\theoremstyle{plain}
\newtheorem{theorem}{Theorem}[section]

\newtheorem{lemma}[theorem]{Lemma}
\newtheorem{proposition}[theorem]{Proposition}
\theoremstyle{definition}
\newtheorem{definition}[theorem]{Definition}
\newtheorem{example}[theorem]{Example}
\newtheorem{remark}[theorem]{Remark}

\def\od#1#2{\dfrac{d#1}{d#2}}
\def\pd#1#2{\dfrac{\partial #1}{\partial #2}}
\def\tpd#1#2{\partial #1/\partial #2}

\def\parentheses#1{\left(#1\right)}
\def\brackets#1{\left[#1\right]}

\def\sech{\mathop{\mathrm{sech}}\nolimits}

\def\DS{\displaystyle}
\def\R{\mathbb{R}}
\def\defeq{\mathrel{\mathop:}=}

\def\setdef#1#2{ \left\{ #1 \ |\ #2 \right\} }
\def\ip#1#2{\left\langle#1,#2\right\rangle}

\long\def\symbolfootnote[#1]#2{\begingroup%
  \def\thefootnote{\fnsymbol{footnote}}\footnote[#1]{#2}\endgroup} 

\def\FL{\mathbb{F}L}

\begin{document}
\footskip=.6in

\maketitle

\begin{abstract}
  We discuss an extension of the Hamilton--Jacobi theory to nonholonomic mechanics with a particular interest in its application to exactly integrating the equations of motion.
  We give an intrinsic proof of a nonholonomic analogue of the Hamilton--Jacobi theorem.
  Our intrinsic proof clarifies the difference from the conventional Hamilton--Jacobi theory for unconstrained systems.
  The proof also helps us identify a geometric meaning of the conditions on the solutions of the Hamilton--Jacobi equation that arise from nonholonomic constraints.
  The major advantage of our result is that it provides us with a method of integrating the equations of motion just as the unconstrained Hamilton--Jacobi theory does.
  In particular, we build on the work by \citet*{IgLeMa2008} so that the conventional method of separation of variables applies to some nonholonomic mechanical systems.
  We also show a way to apply our result to systems to which separation of variables does not apply.
\end{abstract}

\section{Introduction}
\subsection{The Hamilton--Jacobi Theory}
The Hamilton--Jacobi theory for unconstrained systems is well understood both from the classical and geometric points of view.
Besides its fundamental aspects, such as its relation to the action integral and generating functions of symplectic maps, the theory is known to be very useful in exactly integrating Hamilton's equations using the technique of separation of variables~\citep[see, e.g.,][]{GoPoSa2001, Ar1991, La1986}. See also \citet[][Chapter~5]{AbMa1978} for an elegant geometric treatment of the Hamilton--Jacobi theory.

\subsection{Extension to Nonholonomic Mechanics}
Our objective is to extend the Hamilton--Jacobi theory to nonholonomic systems, that is, mechanical systems with non-integrable velocity constraints, building on the previous work by \citet{IgLeMa2008}.
Nonholonomic mechanics deals with such systems by extending the ideas of Lagrangian and Hamiltonian mechanics~\citep[see, e.g.,][]{Bl2003}.
However it is often not straightforward to extend the ideas of unconstrained dynamics to nonholonomic systems, since a mechanical system loses some properties that are common to (conventional) Lagrangian and Hamiltonian systems when one adds nonholonomic constraints.

Since the Hamilton--Jacobi theory is developed based on the Hamiltonian picture of dynamics, a natural starting point in extending the Hamilton--Jacobi theory to nonholonomic systems is a Hamiltonian formulation of nonholonomic mechanics.
\citet{BaSn1993} and \citet{VaMa1994} generalized the definition of Hamiltonian system to the almost-symplectic and almost-Poisson formulations, respectively~\citep[see also][]{KoMa1997c, KoMa1998, Bl2003}.
As is shown in these papers, adding nonholonomic constraints to a Hamiltonian system renders the flow of the system non-symplectic.
In fact, \citet{VaMa1994} showed that the condition for the almost-Poisson Hamiltonian system to be (strictly) Poisson is equivalent to the system being holonomic.
This implies that the conventional Hamilton--Jacobi theory does not directly apply to nonholonomic mechanics, since the (strict) symplecticity is critical in the theory.
In fact, the Hamilton--Jacobi equation is a PDE for generating functions that yield symplectic maps for the flows of the dynamics.

There are some previous attempts to extend the Hamilton--Jacobi theory to nonholonomic mechanics, such as \citet{Pa2005}.
However, as pointed out by \citet{IgLeMa2008}, these results are based on a variational approach, which does not apply to nonholonomic setting.
See \citet{LeMaMa2009} for details.

\citet{IgLeMa2008} proved a nonholonomic Hamilton--Jacobi theorem that shares the geometric view with the unconstrained theory by \citet{AbMa1978}.
The recent work by \citet{LeMaMa2009} developed a new geometric framework for systems defined with linear almost Poisson structures.
Their result generalizes the Hamilton--Jacobi theory to the linear almost Poisson settings, and also specializes and provides geometric insights into nonholonomic mechanics.

\subsection{Nonholonomic Hamilton--Jacobi Theory}
\label{sec:intro-NHJT}
The previous work by \citet{IgLeMa2008} and \citet{LeMaMa2009} is of theoretical importance in its own right.
However, it is still unknown if the theorems are applicable to the problem of exactly integrating the equations of motion of nonholonomic systems in a similar way to the conventional theory.
To see this let us briefly discuss the difference between the unconstrained Hamilton--Jacobi equation and the nonholonomic ones mentioned above.
First recall the conventional unconstrained theory: Let $Q$ be a configuration space, $T^{*}Q$ be its cotangent bundle, and $H: T^{*}Q \to \R$ be the Hamiltonian; then the Hamilton--Jacobi equation can be written as a {\em single} equation:
\begin{subequations}
  \label{eq:HJEq}
  \begin{equation}
    H\parentheses{ q, \pd{W}{q} } = E,
  \end{equation}
  or
  \begin{equation}
    H \circ dW(q) = E,
  \end{equation}
\end{subequations}
for an unknown {\em function} $W: Q \to \R$.
On the other hand, the nonholonomic Hamilton--Jacobi equations in \cite{IgLeMa2008} have the following form:
\begin{equation}
  \label{eq:NHHJEq-Es}
  d(H \circ \gamma)(q) \in \mathcal{D}^{\circ},
\end{equation}
where $\gamma: Q \to T^{*}Q$ is an unknown {\em one-form}, and $\mathcal{D}^{\circ}$ is the annihilator of the distribution $\mathcal{D} \subset TQ$ defined by the nonholonomic constraints.
While it is clear that Eq.~\eqref{eq:NHHJEq-Es} reduces to Eq.~\eqref{eq:HJEq} for the special case that there are no constraints\footnote{$\mathcal{D} = TQ$ and hence $\mathcal{D}^{\circ} = 0$ and identifying the one-form $\gamma$ with $dW$}, Eq.~\eqref{eq:NHHJEq-Es} in general gives a set of partial differential equations for $\gamma$ as opposed to a single equation like Eq.~\eqref{eq:HJEq}.

Having this difference in mind, let us now consider the following question: Is separation of variables applicable to the nonholonomic Hamilton--Jacobi equation?
First recall how separation of variables works in the conventional setting: One first assumes that the function $W$ can be split into pieces, each of which depends only on some subset of the variables $q$, e.g.,
\begin{equation*}
  W(q) = W_{1}(q_{1}) + W_{2}(q_{2}),
\end{equation*}
for $W_{1}, W_{2}: Q \to \R$, and $q = (q_{1}, q_{2})$.
Then this sometimes helps us split the left-hand side of the Hamilton--Jacobi equation \eqref{eq:HJEq}:
\begin{equation*}
  H_{1}\parentheses{ q_{1}, \pd{W_{1}}{q_{1}} } + H_{2}\parentheses{ q_{2}, \pd{W_{2}}{q_{2}} } = E,
\end{equation*}
with some functions $H_{1}, H_{2}: T^{*}Q \to \R$, and hence both $H_{1}$ and $H_{2}$ must be constant:
\begin{equation*}
  H_{1}\parentheses{ q_{1}, \pd{W_{1}}{q_{1}} } = E_{1},
  \qquad
  H_{2}\parentheses{ q_{2}, \pd{W_{2}}{q_{2}} } = E_{2},
\end{equation*}
where $E_{1}$ and $E_{2}$ are constants such that $E_{1} + E_{2} = E$.
Then we can solve them to obtain $\tpd{W_{1}}{q_{1}}$ and $\tpd{W_{2}}{q_{2}}$ separately.
It is not clear how this approach applies to the nonholonomic Hamilton--Jacobi Equation \eqref{eq:NHHJEq-Es}.
Furthermore, there are additional conditions on the solution $\gamma$ which do not exist in the conventional theory.

\subsection{Integrability of Nonholonomic Systems}
Integrability of Hamiltonian systems is an interesting question that has a close link with the Hamilton--Jacobi theory.
For unconstrained Hamiltonian systems, the Arnold--Liouville theorem~\citep[see, e.g.,][]{Ar1991} stands as the definitive work.
The link between the theorem and the Hamilton--Jacobi theory lies in the action-angle variables, which specify the natural canonical coordinates for the invariant tori of the system; in practice the action-angle variables can be found through separation of variables for the Hamilton--Jacobi equation~\citep[see, e.g.,][\S 6.2]{JoSa1998}.

For nonholonomic mechanics, however, the Arnold--Liouville theorem does not directly apply, since the nonholonomic flow is not Hamiltonian and so the key ideas in the Arnold--Liouville theorem lose their effectiveness.
\citet{Ko2002} gave certain conditions for integrability of nonholonomic systems with invariant measures.
However, it is important to remark that there are examples that do not have invariant measures but are still integrable, such as the Chaplygin sleigh~\citep[see, e.g.,][]{Bl2000, Bl2003}.
Also it is unknown how this result may be related to the nonholonomic Hamilton--Jacobi theory, which does not have an apparent relationship with invariant measures.

\subsection{Main Results}
The goal of the present paper is to fill the gap between the unconstrained and nonholonomic Hamilton--Jacobi theory by showing applicability of separation of variables to nonholonomic systems, and also to discuss integrability of them.
For that purpose, we would like to first reformulate the nonholonomic Hamilton--Jacobi theorem from an intrinsic point of view\footnote{A coordinate-based proof is given in \cite{IgLeMa2008}}.
We show that the nonholonomic Hamilton--Jacobi equation \eqref{eq:NHHJEq-Es} reduces to a single equation $H \circ \gamma = E$.
This result resolves the differences between unconstrained and nonholonomic Hamilton--Jacobi equations mentioned in Section~\ref{sec:intro-NHJT}, and makes it possible to apply separation of variables to nonholonomic systems.
Furthermore, the intrinsic proof helps us identify the difference from the unconstrained theory by \citet{AbMa1978} and find the conditions on the solution $\gamma$ arising from nonholonomic constraints that are more practical than (although equivalent to, as pointed out by \citet{So2009}) those of \citet{IgLeMa2008}.
It turns out that these conditions are not only useful in finding the solutions of the Hamilton--Jacobi equation by separation of variables, but also provide a way to integrate the equations of motion of a system to which separation of variables does not apply.

\subsection{Outline of the Paper}
In Section~\ref{sec:HFNM} we briefly review the Hamiltonian formulation of nonholonomic mechanics, and also state some definitions and results that pertain to the nonholonomic Hamilton--Jacobi theorem.
In particular, we first give an intrinsic description of nonholonomic Hamilton equations, define and state a few results concerning completely nonholonomic constraints and regularity of nonholonomic systems.
Much of the ideas in the proof of the nonholonomic Hamilton--Jacobi theorem come from identifying both the similarities and differences between the nonholonomic and unconstrained Hamilton equations.

In Section~\ref{sec:NHHJThm} we formulate and prove the nonholonomic Hamilton--Jacobi theorem.
The theorem and proof are an extension of the one by \citet{AbMa1978} to the nonholonomic setting.
In doing so we identify the differences from the unconstrained theory; this in turn gives the additional conditions arising from the nonholonomic constraints.

We apply the nonholonomic Hamilton--Jacobi theorem to several examples in Section~\ref{sec:App}.
We first apply the technique of separation of variables to solve the nonholonomic Hamilton--Jacobi equation to obtain exact solutions of the motions of the vertical rolling disk and knife edge on an inclined plane.
We then take the snakeboard and Chaplygin sleigh as examples to which separation of variables does not apply, and show another way of employing the nonholonomic Hamilton--Jacobi theorem to exactly integrate the equations of motion.
The conclusion follows to suggest possible future work.

\section{Hamiltonian Formulation of Nonholonomic Mechanics}
\label{sec:HFNM}
\subsection{Hamilton's Equations for Nonholonomic Systems}
Hamiltonian approaches to nonholonomic mechanical systems are developed by, for example, \citet{BaSn1993} and \citet{VaMa1994}. See also \citet{KoMa1997c, KoMa1998} and \citet{Bl2003}.

Consider a mechanical system on a differentiable manifold $Q$ with Lagrangian $L: TQ \to \R$.
Suppose that the system has nonholonomic constraints given by the distribution
\begin{equation}
  \label{eq:D}
  \mathcal{D} \defeq \setdef{ v \in TQ }{ \omega^{s}(v) = A_{i}^{s} v^{i} = 0,\, s = 1,\dots,p },
\end{equation}
where $\lambda_{s}$ are Lagrange multipliers and $\omega^{s} = A_{i}^{s}\,dq^{i}$ are linearly independent non-exact one-forms on $Q$.
Then the Lagrange--d'Alembert principle gives the equation of motion \citep[see, e.g.][Chapter~5]{Bl2003}:
\begin{equation}
  \label{eq:LdA}
  \od{}{t}\pd{L}{\dot{q}^{i}} - \pd{L}{q^{i}} = \lambda_{s} A^{s}_{i}.
\end{equation}
The Legendre transformation of this set of equations gives the Hamiltonian formulation of nonholonomic systems.
Specifically, define the Legendre transform $\FL: TQ \to T^{*}Q$ by
\begin{equation*}
  \FL(v_{q}) \cdot w_{q} = \left.\od{}{\varepsilon} L(v_{q} + \varepsilon\,w_{q}) \right|_{\varepsilon=0},
\end{equation*}
for $v_{q}, w_{q} \in T_{q}Q$.
Throughout the paper we assume that the Lagrangian is hyperregular, i.e., the Legendre transform $\FL$ is a diffeomorphism.
Set $p \defeq \FL(\dot{q})$, or locally $p_{i} = \tpd{L}{\dot{q}^{i}}$, and define the Hamiltonian $H: T^{*}Q \to \R$ by
\begin{equation*}
  H(q,p) \defeq \ip{p}{\dot{q}} - L(q,\dot{q}),
\end{equation*}
where $\dot{q} = (\FL)^{-1}(p)$ on the right-hand side.
Then we can rewrite Eq.~\eqref{eq:LdA} as follows:
\begin{equation}
  \label{eq:NHHam-local}
  \dot{q}^{i} = \pd{H}{p_{i}},
  \qquad
  \dot{p_{i}} = -\pd{H}{q^{i}} + \lambda_{s} A^{s}_{i},
\end{equation}
with the constraint equations
\begin{equation}
  \label{eq:NHHam-condition}
  \omega^{s}(\dot{q}) = \omega^{s}\parentheses{ \pd{H}{p} } = 0
  \quad\text{for}\quad
  s = 1, \dots, p.
\end{equation}
Equations~\eqref{eq:NHHam-local} and \eqref{eq:NHHam-condition} define {\em Hamilton's equations for nonholonomic systems}.
We can also write this system in the intrinsic form in the following way:
Suppose that $X_{H}^{\rm nh} = \dot{q}^{i}\partial_{q^{i}} + \dot{p}_{i}\partial_{p_{i}}$ is the vector field on $T^{*}Q$ that defines the flow of the system, $\Omega$ is the standard symplectic form on $T^{*}Q$, and $\pi_{Q}: T^{*}Q \to Q$ is the cotangent bundle projection.
Then we can write Hamilton's equations for nonholonomic systems~\eqref{eq:NHHam-local} and \eqref{eq:NHHam-condition} in the following intrinsic form:
\begin{equation}
  \label{eq:NHHam-intrinsic}
  i_{X_{H}^{\rm nh}} \Omega = dH - \lambda_{s} \pi_{Q}^{*} \omega^{s},
\end{equation}
along with
\begin{equation}
  \label{eq:NHHam-condition-intrinsic}
  T\pi_{Q}(X_{H}^{\rm nh}) \in \mathcal{D}
  \quad\text{or}\quad
  \omega^{s}(T\pi_{Q}(X_{H}^{\rm nh})) = 0
  \ \text{for}\ s = 1, \dots, p.
\end{equation}
Introducing the {\em constrained momentum space} $\mathcal{M} \defeq \FL(\mathcal{D}) \subset T^{*}Q$, the above constraints may be replaced by the following:
\begin{equation}
  \label{eq:NHHam-condition2-intrinsic}
  p \in \mathcal{M}.
\end{equation}

\subsection{Completely Nonholonomic Constraints}
Let us introduce a special class of nonholonomic constraints that is assumed in the nonholonomic Hamilton--Jacobi theorem\footnote{We would like to thank the referees for pointing out the importance of the notion.}.
\begin{definition}[\citet{VeGe1988}; see also \citet{Mo2002}]
  \label{def:CompletelyNonholonomic}
  A distribution $\mathcal{D} \subset TQ$ is said to be {\em completely nonholonomic} (or {\em bracket-generating}) if $\mathcal{D}$ along with all of its iterated Lie brackets $[\mathcal{D}, \mathcal{D}], [\mathcal{D}, [\mathcal{D}, \mathcal{D}]], \dots$ spans the tangent bundle $TQ$.
\end{definition}
Let us also introduce the following notion for convenience:
\begin{definition}
  Let $Q$ be the configuration manifold of a mechanical system.
  Then nonholonomic constraints on the system are said to be {\em completely nonholonomic} if the distribution $\mathcal{D} \subset TQ$ defined by the nonholonomic constraints is completely nonholonomic (or bracket-generating).
\end{definition}
One of the most important results concerning completely nonholonomic distributions is the following~\footnote{See, e.g., \citet{Mo2002} for a proof.}:
\begin{theorem}[Chow's Theorem]
  Let $Q$ be a connected differentiable manifold.
  If a distribution $\mathcal{D} \subset TQ$ is completely nonholonomic, then any two points on $Q$ can be joined by a horizontal path.
\end{theorem}
We will need the following result that easily follows from Chow's Theorem:
\begin{proposition}
  \label{prop:exact_one-forms_and_CND}
  Let $Q$ be a connected differentiable manifold and $\mathcal{D} \subset TQ$ be a completely nonholonomic distribution.
  Then there is no non-zero exact one-form in the annihilator $\mathcal{D}^{\circ} \subset T^{*}Q$.
\end{proposition}
\begin{proof}
  Chow's Theorem says that, for any two points $q_{0}$ and $q_{1}$ in $Q$, there exists a curve $c: [0,T] \to Q$ with some $T > 0$ such that $c(0) = q_{0}$ and $c(T) = q_{1}$, and also $\dot{c}(t) \in \mathcal{D}_{c(t)}$ for any $t \in (0,T)$.
  Now let $df$ be an exact one-form in the annihilator $\mathcal{D}^{\circ}$.
  Then by Stokes' theorem, we have
  \begin{equation*}
    f(q_{1}) - f(q_{0}) = \int_{0}^{T} df( \dot{c}(t) )\,dt = 0,
  \end{equation*}
  where $df( \dot{c}(t) ) = 0$ because $df \in \mathcal{D}^{\circ}$ and $\dot{c}(t) \in \mathcal{D}_{c(t)}$.
  Since $q_{0}$ and $q_{1}$ are arbitrary and $Q$ is connected, this implies that $f$ is constant on $Q$.
\end{proof}

\subsection{Regularity of Nonholonomic Systems}
We will also need to assume {\em regularity} of nonholonomic systems in the following sense\footnote{We again would like to thank one of the referees for pointing out the necessity of this assumption.}:
Consider a nonholonomic system with a hyperregular Lagrangian $L: TQ \to \R$ and a constant-dimensional distribution $\mathcal{D} \subset TQ$ defined by nonholonomic constraints.
For any $v_{q} \in TQ$ define a bilinear form $B_{L}(v_{q}): T_{q}Q \times T_{q}Q \to \R$ by
\begin{equation*}
  B_{L}(v_{q})(u_{q}, w_{q}) \defeq \left.\pd{^{2}}{\varepsilon_{1} \partial\varepsilon_{2}} L(v_{q} + \varepsilon_{1} u_{q} + \varepsilon_{2} w_{q}) \right|_{\varepsilon_{1} = \varepsilon_{2} = 0} = D_{2}D_{2}L(q, v) \cdot (u_{q}, w_{q}).
\end{equation*}
Then hyperregularity of the Lagrangian implies that the associated map $B_{L}^{\flat}(v_{q}): T_{q}Q \to T_{q}^{*}Q$ defined by
\begin{equation*}
  \ip{ B_{L}^{\flat}(v_{q})(u_{q}) }{ w_{q} } \defeq B_{L}(v_{q})(u_{q}, w_{q})
\end{equation*}
is an isomorphism.
Thus we can define a bilinear form $W_{L}: T_{q}^{*}Q \times T_{q}^{*}Q \to \R$ by
\begin{equation*}
  W_{L}(v_{q})(\alpha_{q}, \beta_{q}) \defeq \ip{ \alpha_{q} }{ (B_{L}^{\flat})^{-1}(\beta_{q}) }.
\end{equation*}

\begin{definition}[\citet{LeMa1996e}; see also \citet{LeMaDi1997}]
  \label{def:Regularity}
  In the above setup, suppose that the annihilator $\mathcal{D}^{\circ}$ is spanned by the one-forms $\{ \omega^{s} \}_{s = 1}^{p}$.
  Then the nonholonomic system is said to be {\em regular} if the matrices $(\mathcal{C}_{L}^{rs}(v))$ defined by
  \begin{equation}
    \label{eq:mathcalC}
    \mathcal{C}_{L}^{rs}(v) = -W_{L}(v)( \omega^{r}, \omega^{s} )
  \end{equation}
  are nonsingular for any $v \in \mathcal{D}$.
\end{definition}

For a mechanical system whose Lagrangian is kinetic minus potential energy, regularity follows automatically:
\begin{proposition}[\citet{CaRa1993}; see also \citet{LeMa1996e}]
  If the Lagrangian $L: TQ \to \R$ has the form
  \begin{equation}
    \label{eq:SimpleLagrangian}
    L(q,v) = \frac{1}{2}g(v,v) - V(q),
  \end{equation}
  with $g$ being a Riemannian metric on $Q$, then the nonholonomic system is regular.
\end{proposition}
\begin{proof}
  In this case $D_{2}D_{2}L(q,v)(u_{q}, w_{q}) = g_{q}(u_{q}, w_{q})$, and so $W_{L}$ is defined by the inverse $g^{ij}$ of the matrix $g_{ij}$.
  Since $g_{ij}$ is positive-definite, so is the inverse $g^{ij}$; hence it follows that $W_{L}$ is positive-definite.
  A positive-definite matrix restricted to a subspace is again positive-definite, and so $\mathcal{C}_{L}^{rs}$ is positive-definite and hence nondegenerate.
\end{proof}

In the Hamiltonian setting with the form of Lagrangian in Eq.~\eqref{eq:SimpleLagrangian}, we have the following result:
\begin{proposition}[\citet{BaSn1993}]
  \label{prop:BaSn1993}
  Suppose that the Lagrangian is of the form in Eq.~\eqref{eq:SimpleLagrangian}.
  Let $\mathcal{F}$ be the distribution on $T^{*}Q$ defined by
  \begin{equation}
    \label{eq:mathcalF}
    \mathcal{F} \defeq \setdef{ v \in TT^{*}Q }{ T\pi_{Q}(v) \in \mathcal{D} },
  \end{equation}
  and then define a distribution $\mathcal{H}$ on $\mathcal{M} \defeq \FL(\mathcal{D})$ by
  \begin{equation}
    \label{eq:mathcalH}
    \mathcal{H} \defeq \mathcal{F} \cap T\mathcal{M}.
  \end{equation}
  Then the standard symplectic form $\Omega$ restricted to $\mathcal{H}$ is nondegenerate. 
\end{proposition}
\begin{proof}
  See \citet[][Theorem on p.~105]{BaSn1993}.
\end{proof}

\section{Nonholonomic Hamilton--Jacobi Theorem}
\label{sec:NHHJThm}
We would like to refine the result of \citet{IgLeMa2008} with a particular attention to applications to exact integration of the equations of motion.
Specifically, we would like to take an intrinsic approach (see \cite{IgLeMa2008} for the coordinate-based approach) to clarify the difference from the (unconstrained) Hamilton--Jacobi theorem of \citet{AbMa1978} (Theorem~5.2.4).
  A significant difference from the result by \citet{IgLeMa2008} is that the nonholonomic Hamilton--Jacobi equation is given as a single algebraic equation $H \circ \gamma = E$ just as in the unconstrained Hamilton--Jacobi theory, as opposed to a set of differential equations $d(H \circ \gamma) \in \mathcal{D}^{\circ}$.

\begin{theorem}[Nonholonomic Hamilton--Jacobi]
  \label{thm:NHHJ}
  Consider a nonholonomic system defined on a connected differentiable manifold $Q$ with a Lagrangian of the form Eq.~\eqref{eq:SimpleLagrangian} and a completely nonholonomic constraint distribution $\mathcal{D} \subset TQ$.
  Let $\gamma: Q \to T^{*}Q$ be a one-form that satisfies
  \begin{equation}
\label{eq:gamma_in_mathcalM}
    \gamma(q) \in \mathcal{M}_{q} \text{ for any } q \in Q,
  \end{equation}
  and
  \begin{equation}
    d\gamma|_{\mathcal{D}\times\mathcal{D}} = 0, \text{ i.e., } d\gamma(v,w) = 0 \text{ for any } v, w \in \mathcal{D}.
    \label{eq:dgamma}
  \end{equation}
  Then the following are equivalent:
  \begin{enumerate}
    \renewcommand{\theenumi}{\roman{enumi}}
    \renewcommand{\labelenumi}{(\theenumi)}
  \item \label{enumi:NHHJ-i}
    For every curve $c(t)$ in $Q$ satisfying
    \begin{equation}
      \label{eq:NHHJ-curve}
      \dot{c}(t) = T\pi_{Q} \cdot X_{H}( \gamma \circ c(t) ),
    \end{equation}
    the curve $t \mapsto \gamma \circ c(t)$ is an integral curve of $X_{H}^{\rm nh}$, where $X_{H}$ is the Hamiltonian vector field of the unconstrained system with the same Hamiltonian, i.e., $i_{X_{H}}\Omega = dH$.
    \medskip
  \item \label{enumi:NHHJ-ii}
    The one-form $\gamma$ satisfies the {\em nonholonomic Hamilton--Jacobi equation}:
    \begin{equation}
      \label{eq:NHHJ}
      H \circ \gamma = E,
    \end{equation}
    where $E$ is a constant.
  \end{enumerate}
\end{theorem}

The following lemma, which is a slight modification of Lemma~5.2.5 of \citet{AbMa1978}, is the key to the proof of the above theorem:
\begin{lemma}
  \label{lem:NHHJ}
  For any one-form $\gamma$ on $Q$ that satisfies the condition Eq.~\eqref{eq:dgamma} and any $v, w \in \mathcal{F}$, the following equality holds:
  \begin{equation}
    \Omega( T(\gamma\circ\pi_{Q}) \cdot v, w )
    = \Omega( v, w - T(\gamma\circ\pi_{Q}) \cdot w ).
  \end{equation}
\end{lemma}

\begin{proof}
  Notice first that $v - T(\gamma\circ\pi_{Q}) \cdot v$ is vertical for any $v \in TT^{*}Q$:
  \begin{align*}
    T\pi_{Q} \cdot ( v - T(\gamma\circ\pi_{Q}) \cdot v )
    &= T\pi_{Q}(v) - T(\pi_{Q} \circ \gamma \circ \pi_{Q}) \cdot v
    \\
    &= T\pi_{Q}(v) - T\pi_{Q}(v) = 0,
  \end{align*}
  where we used the relation $\pi_{Q} \circ \gamma \circ \pi_{Q} = \pi_{Q}$.
  Hence
  \begin{equation*}
    \Omega( v - T(\gamma\circ\pi_{Q}) \cdot v, w - T(\gamma\circ\pi_{Q}) \cdot w ) = 0,
  \end{equation*}
  and thus
  \begin{equation*}
    \Omega( T(\gamma\circ\pi_{Q}) \cdot v, w )
    = \Omega(v, w - T(\gamma\circ\pi_{Q}) \cdot w )
    + \Omega( T(\gamma\circ\pi_{Q}) \cdot v, T(\gamma\circ\pi_{Q}) \cdot w ).
  \end{equation*}
  However, the second term on the right-hand side vanishes:
  \begin{equation*}
    \Omega( T(\gamma\circ\pi_{Q}) \cdot v, T(\gamma\circ\pi_{Q}) \cdot w )
    = \gamma^{*}\Omega( T\pi_{Q}(v), T\pi_{Q}(w) )
    = -d\gamma( T\pi_{Q}(v), T\pi_{Q}(w) ) = 0,
  \end{equation*}
  where we used the fact that for any one-form $\beta$ on $Q$, $\beta^{*}\Omega = -d\beta$ with $\beta$ on the left-hand side being regarded as a map $\beta: Q \to T^{*}Q$~\citep[See][Proposition~3.2.11 on p.~179]{AbMa1978}, and the assumption that $d\gamma|_{\mathcal{D}\times\mathcal{D}} = 0$; note that $v, w \in \mathcal{F}$ implies $T\pi_{Q}(v), T\pi_{Q}(w) \in \mathcal{D}$.
\end{proof}

Let us state another lemma:
\begin{lemma}
  \label{lem:NHHJ-2}
  The unconstrained Hamiltonian vector field $X_{H}$ evaluated on the constrained momentum space $\mathcal{M}$ is in the distribution $\mathcal{F}$, i.e.,
  \begin{equation*}
    X_{H}(\alpha_{q}) \in \mathcal{F}_{\alpha_{q}}
    \text{ for any }
    \alpha_{q} \in \mathcal{M}_{q}.
  \end{equation*}
\end{lemma}
\begin{proof}
  We want to show that $T\pi_{Q} (X_{H}(\alpha_{q}))$ is in $\mathcal{D}_{q}$.
  First notice that
  \begin{align*}
    T\pi_{Q} (X_{H}(\alpha_{q})) = \pd{H}{p_{i}}(\alpha_{q})\,\pd{}{q^{i}} = \mathbb{F}H(\alpha_{q}),
  \end{align*}
  where we defined $\mathbb{F}H: T^{*}Q \to TQ$ by
  \begin{equation*}
    \ip{ \beta_{q} }{ \mathbb{F}H(\alpha_{q}) } = \left.\od{}{\varepsilon} H(\alpha_{q} + \varepsilon\,\beta_{q}) \right|_{\varepsilon=0}.
  \end{equation*}
  However, because the Lagrangian $L$ is hyperregular, we have $\mathbb{F}H = (\FL)^{-1}$ and thus
  \begin{align*}
    T\pi_{Q} (X_{H}(\alpha_{q})) =  (\FL)^{-1}(\alpha_{q}).
  \end{align*}
  Now, by the definition of $\mathcal{M}$, $\alpha_{q} \in \mathcal{M}$ implies $\alpha_{q} \in \FL(\mathcal{D}_{q})$, which gives $(\FL)^{-1}(\alpha_{q}) \in \mathcal{D}_{q}$ by the hyperregularity of $L$.
  Hence the claim follows.
\end{proof}

\begin{proof}[Proof of Theorem~\ref{thm:NHHJ}]
  Let us first show that \eqref{enumi:NHHJ-ii} implies \eqref{enumi:NHHJ-i}.
  Assume \eqref{enumi:NHHJ-ii} and let $p(t) \defeq \gamma \circ c(t)$, where $c(t)$ satisfies Eq.~\eqref{eq:NHHJ-curve}.
  Then
  \begin{align}
    \dot{p}(t) &= T\gamma(\dot{c}(t))
    \nonumber\\
    &= T\gamma \cdot T\pi_{Q} \cdot X_{H}( \gamma \circ c(t) )
    \nonumber\\
    &= T( \gamma \circ \pi_{Q}) \cdot X_{H}( \gamma \circ c(t) ).
    \label{eq:pdot}
  \end{align}
  Therefore, using Lemmas~\ref{lem:NHHJ} and \ref{lem:NHHJ-2}, we obtain, for any $w \in \mathcal{F}$,
  \begin{align*}
    \Omega( T(\gamma\circ\pi_{Q}) \cdot X_{H}( p(t) ), w )
    &= \Omega( X_{H}( p(t) ), w - T(\gamma\circ\pi_{Q}) \cdot w )
    \\
    &= \Omega( X_{H}( p(t) ), w )
    - \Omega( X_{H}( p(t) ), T(\gamma\circ\pi_{Q}) \cdot w ).
  \end{align*}
  For the first term on the right-hand side, notice that for any $w \in \mathcal{F}$,
  \begin{equation*}
    \Omega(X_{H}^{\rm nh}, w) = dH \cdot w - \lambda_{s}\pi_{Q}^{*}\omega^{s}(w) = dH \cdot w = \Omega(X_{H}, w).
  \end{equation*}
  Also for the second term,
  \begin{equation*}
    \Omega( X_{H}(p(t)), T(\gamma\circ\pi_{Q}) \cdot w )
    = dH(p(t)) \cdot T(\gamma\circ\pi_{Q}) \cdot w
    = d( H \circ \gamma )(c(t)) \cdot T\pi_{Q}(w).
  \end{equation*}
  So we now have
  \begin{equation}
    \label{eq:NHHJ-rel}
    \Omega( T(\gamma\circ\pi_{Q}) \cdot X_{H}( p(t) ), w )
    = \Omega( X_{H}^{\rm nh}( p(t) ), w )
    - d( H \circ \gamma )(c(t)) \cdot T\pi_{Q}(w).
  \end{equation}
  However, the nonholonomic Hamilton--Jacobi equation \eqref{eq:NHHJ} implies that the second term on the right-hand side vanishes.
  Thus we have
  \begin{equation}
    \label{eq:X_H-X_H^nh}
    \Omega( T(\gamma\circ\pi_{Q}) \cdot X_{H}( p(t) ), w )
    = \Omega( X_{H}^{\rm nh}( p(t) ), w )
  \end{equation}
  for any $w \in \mathcal{F}_{p(t)}$.
  Now $T(\gamma\circ\pi_{Q}) \cdot X_{H} \in T\mathcal{M}$ since $\gamma$ takes values in $\mathcal{M}$; also $T(\gamma\circ\pi_{Q}) \cdot X_{H}( p(t) ) \in \mathcal{F}_{p(t)}$ because
  \begin{equation*}
    T\pi_{Q} \cdot T(\gamma\circ\pi_{Q}) \cdot X_{H}( p(t) )
    = T(\pi_{Q} \circ \gamma \circ \pi_{Q}) \cdot X_{H}( p(t) )
    = T\pi_{Q} \cdot X_{H}( p(t) ) \in \mathcal{D},
  \end{equation*}
  using Lemma~\ref{lem:NHHJ-2} again.
  Therefore $T(\gamma\circ\pi_{Q}) \cdot X_{H}( p(t) ) \in \mathcal{H}_{p(t)}$.
  On the other hand, $X_{H}^{\rm nh}( p(t) ) \in \mathcal{H}_{p(t)}$ as well: $X_{H}^{\rm nh}( p(t) ) \in T_{p(t)}\mathcal{M}$ because $\mathcal{M}$ is an invariant manifold of the nonholonomic flow defined by $X_{H}^{\rm nh}$ and also $X_{H}^{\rm nh}( p(t) ) \in \mathcal{F}_{p(t)}$ due to Eq.~\eqref{eq:NHHam-condition-intrinsic}.
  Now, in Eq.~\eqref{eq:X_H-X_H^nh}, $w$ is an arbitrary element in $\mathcal{F}_{p(t)}$ and thus Eq.~\eqref{eq:X_H-X_H^nh} holds for any $w \in \mathcal{H}_{p(t)}$ because $\mathcal{H} \subset \mathcal{F}$.
  However, according to Proposition~\ref{prop:BaSn1993}, $\Omega$ restricted to $\mathcal{H}$ is nondegenerate.
  So we obtain
  \begin{equation*}
    T(\gamma\circ\pi_{Q}) \cdot X_{H}( p(t) ) = X_{H}^{\rm nh}( p(t) ),
  \end{equation*}
  and hence Eq.~\eqref{eq:pdot} gives
  \begin{equation*}
    \dot{p}(t) = X_{H}^{\rm nh}( p(t) ).
  \end{equation*}
  This means that $p(t)$ gives an integral curve of $X_{H}^{\rm nh}$.
  Thus \eqref{enumi:NHHJ-ii} implies \eqref{enumi:NHHJ-i}.

  Conversely, assume \eqref{enumi:NHHJ-i}; let $c(t)$ be a curve in $Q$ that satisfies Eq.~\eqref{eq:NHHJ-curve} and set $p(t) \defeq \gamma \circ c(t)$.
  Then $p(t)$ is an integral curve of $X_{H}^{\rm nh}$ and so
  \begin{equation*}
    \dot{p}(t) = X_{H}^{\rm nh}(p(t)).
  \end{equation*}
  However, from the definition of $p(t)$ and Eq.~\eqref{eq:NHHJ-curve},
  \begin{equation*}
    \dot{p}(t) = T\gamma( \dot{c}(t) ) = T\gamma \cdot T\pi_{Q} \cdot X_{H}( p(t) )
     = T(\gamma \circ \pi_{Q}) \cdot X_{H}( p(t) ).
  \end{equation*}
  Therefore we get 
  \begin{equation*}
    X_{H}^{\rm nh}(p(t)) = T(\gamma \circ \pi_{Q}) \cdot X_{H}( p(t) ).
  \end{equation*}
  In view of Eq.~\eqref{eq:NHHJ-rel}, we get, for any $w \in TT^{*}Q$ such that $T\pi_{Q}(w) \in \mathcal{D}$,
  \begin{equation*}
    d( H \circ \gamma )(c(t)) \cdot T\pi_{Q}(w) = 0,
  \end{equation*}
  but this implies $d( H \circ \gamma )(c(t)) \cdot v = 0$ for any $v \in \mathcal{D}_{c(t)}$, or $d( H \circ \gamma )(c(t)) \in \mathcal{D}_{c(t)}^{\circ}$.
  However, this further implies $d(H \circ \gamma)(q) = 0$ for any $q \in Q$: For an arbitrary point $q \in Q$, consider a curve $c(t)$ that satisfies Eq.~\eqref{eq:NHHJ-curve} such that $c(0) = q$.
  Then this gives $d(H \circ \gamma)(q) \in \mathcal{D}^{\circ}_{q}$.
  Therefore $d(H \circ \gamma) \in \mathcal{D}^{\circ}$ on $Q$, but then Proposition~\ref{prop:exact_one-forms_and_CND} implies that $d(H \circ \gamma) = 0$ because $\mathcal{D}$ is assumed to be completely nonholonomic.
  Therefore we have $H \circ \gamma = E$ for some constant $E$, which is the nonholonomic Hamilton--Jacobi equation~\eqref{eq:NHHJ}.
\end{proof}

\begin{remark}
  \label{rem:dgamma-1}
  The condition on $d\gamma$, Eq.~\eqref{eq:dgamma}, stated in the above theorem is equivalent to the one in \cite{IgLeMa2008} as pointed out by \citet{So2009}~\citep[see also][Lemma~4.6 on p.~51]{Mo2002}.
  However Eq.~\eqref{eq:dgamma} gives a simpler geometric interpretation and also is easily implemented in applications.
  To be specific, the condition in \cite{IgLeMa2008} states that there exist one-forms $\{ \beta^{i} \}_{i=1}^{p}$ such that 
  \begin{equation}
    d\gamma = \sum_{s=1}^{p} \beta^{s} \wedge \omega^{s},
  \end{equation}
  which does not easily translate into direct expressions for the conditions on $\gamma$.
  On the other hand, Eq.~\eqref{eq:dgamma} is equivalent to
  \begin{equation}
    d\gamma(v_{i}, v_{j}) = 0 \text{ for any } i \neq j,
  \end{equation}
  where $\{v_{i}\}_{i=1}^{n-p}$ spans the distribution $\mathcal{D}$.
  Clearly the above equations give direct expressions for the conditions on $\gamma$.
  We will see later in Section~\ref{sec:App} that the above equations play an important role in exact integration.
\end{remark}

\begin{remark}
  \label{rem:dgamma-2}
  Table~\ref{tab:ComparisonBtwUnconstNHHJ} compares Theorem~\ref{thm:NHHJ} with the unconstrained Hamilton--Jacobi theorem of \citet{AbMa1978} (Theorem~5.2.4).
  Note that Eq.~\eqref{eq:dgamma} is trivially satisfied for the unconstrained case.
  Recall that $\gamma$ is replaced by an exact one-form $dW$ in this case.
  Since $\mathcal{D} = TQ$ by assumption, we have $d\gamma|_{\mathcal{D}\times\mathcal{D}} = d\gamma = d(dW) = 0$ and thus this does not impose any condition on $dW$.
  \begin{table}[htbp]
    \centering
    \caption{Comparison between unconstrained and nonholonomic Hamilton--Jacobi theorems.}
    \label{tab:ComparisonBtwUnconstNHHJ}
    \small
    \begin{tabular}{|c||c|c|}
      \hline
      & {\bfseries\sffamily Unconstrained}\bigstrut & {\bfseries\sffamily Nonholonomic}\bigstrut
      \\\hline\hline
      {\sf Generating Function} & $\bigstrut W: Q \to \mathbb{R}$ & None
      \\\hline
      {\sf One-form} & $\bigstrut dW: Q \to T^{*}Q$ & $\bigstrut \gamma: Q \to \mathcal{M} \subset T^{*}Q$
      \\\hline
      {\sf Condition} & $\bigstrut ddW = 0$~(trivial) & $\bigstrut d\gamma|_{\mathcal{D} \times \mathcal{D}} = 0$
      \\\hline
      \multirow{2}{*}{\sf Hamilton--Jacobi Eq.}
      &
      \multirow{2}{*}{$\displaystyle H \circ dW(q) = E$}
      &
      \multirow{2}{*}{$\displaystyle H \circ \gamma(q) = E$}
      \\
      & &
      \\\hline
    \end{tabular}
  \end{table}
\end{remark}

\begin{remark}
  See \citet{CaGrMaMaMuRo2009} for a Lagrangian version of Theorem~\ref{thm:NHHJ}, and \citet{LeMaMa2009} for an extension to a more general framework, i.e., systems defined with linear almost Poisson structures.
\end{remark}

\section{Application to Exactly Integrating Equations of Motion}
\label{sec:App}
\subsection{Applying the Nonholonomic Hamilton--Jacobi Theorem to Exact Integration}
\label{ssec:AHJTEI}
\begin{figure}[hbtp]
  \centering
  \includegraphics[width=.5\linewidth]{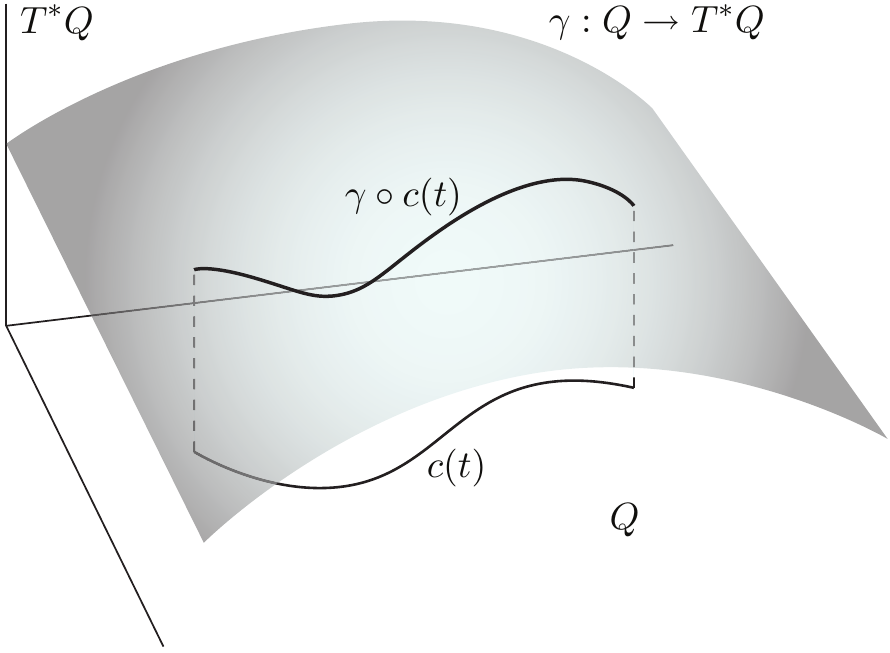}
  \caption{Schematic of an implication of the nonholonomic Hamilton--Jacobi theorem.}
  \label{fig:NHHJ}
\end{figure}

Theorem~\ref{thm:NHHJ} suggests a way to use the solution of the Hamilton--Jacobi equation to integrate the equations of motion.
Namely,
\begin{enumerate}
  \renewcommand{\theenumi}{\arabic{enumi}}
  \renewcommand{\labelenumi}{\sf Step~\theenumi.}
\item Find a solution $\gamma(q)$ of the Hamilton--Jacobi
  equation
  \begin{equation}
    H \circ \gamma (q) = E,
  \end{equation}
  that satisfies the conditions $\gamma(q) \in \mathcal{M}_{q}$ and $d\gamma|_{\mathcal{D} \times \mathcal{D}} = 0$;
  \smallskip
\item Substitute the solution $\gamma(q)$ into Eq.~\eqref{eq:NHHJ-curve} to obtain the set of first-order ODEs defined in the configuration $Q$:
  \begin{subequations}
    \label{eq:sec-App:NHHJ-curve}
    \begin{equation}
      \dot{c}(t) = T\pi_{Q} \cdot X_{H}( \gamma \circ c(t) ),
    \end{equation}
    or, in coordinates,
    \begin{equation}
      \dot{c}(t) = \pd{H}{p}( \gamma \circ c(t) );
    \end{equation}
  \end{subequations}
  \smallskip
\item Solve the ODEs \eqref{eq:sec-App:NHHJ-curve} to find the curve $c(t)$ in the configuration space $Q$.
  Then $\gamma \circ c(t)$ gives the dynamics in the phase space $T^{*}Q$.
\end{enumerate}
Figure~\ref{fig:NHHJ} depicts the idea of this procedure.

In the following sections, we apply this procedure to several examples of nonholonomic systems.
In any of the examples to follow, it is easy to check that the constraints are completely nonholonomic~(see Definition~\ref{def:CompletelyNonholonomic}), and also that the Lagrangian takes the form in Eq.~\eqref{eq:SimpleLagrangian} and hence the system is regular in the sense of Definition~\ref{def:Regularity}.

\subsection{Examples with Separation of Variables}
\label{ssec:WSOV}
Let us first illustrate through a very simple example how the above procedure works with the method of separation of variables.

\begin{example}[The vertical rolling disk]
  \label{ex:VRD}
  \citep[See, e.g.,][]{Bl2003}.
  Consider the motion of the vertical rolling disk of radius $R$ shown in Fig.~\ref{fig:VRD}.
  \begin{figure}[htbp]
    \centering
    \includegraphics[width=.65\linewidth]{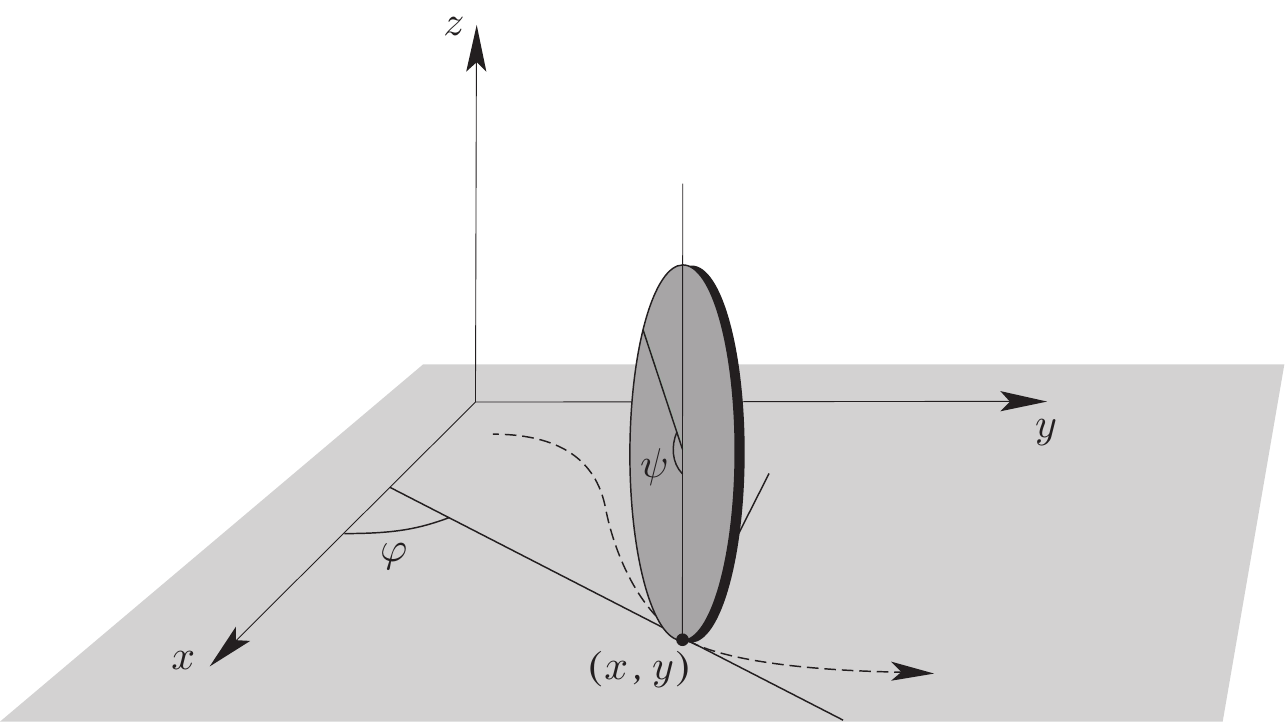}
    \caption{Vertical rolling disk.}
    \label{fig:VRD}
  \end{figure}
  The configuration space is $Q = SE(2) \times S^{1} = \{ (x, y, \varphi, \psi) \}$.
  Suppose that $m$ is the mass of the disk, $I$ is the moment of inertia of the disk about the axis perpendicular to the plane of the disk, and $J$ is the moment of inertia about an axis in the plane of the disk (both axes passing through the disk's center).
  The velocity constraints are 
  \begin{equation}
    \label{eq:constraints-VRD}
    \dot{x} = R\cos\varphi\,\dot{\psi},
    \qquad
    \dot{y} = R\sin\varphi\,\dot{\psi},
  \end{equation}
  or in terms of constraint one-forms, 
  \begin{equation}
    \label{eq:omegas-VRD}
    \omega^{1} = dx - R\cos\varphi\,d\psi,
    \qquad
    \omega^{2} = dy - R\sin\varphi\,d\psi.
  \end{equation}
  The Hamiltonian $H: T^{*}Q \to \R$ is given by
  \begin{equation}
    H = \frac{1}{2}\parentheses{ \frac{p_{x}^{2} + p_{y}^{2}}{m} + \frac{p_{\varphi}^{2}}{J} + \frac{p_{\psi}^{2}}{I} }.
  \end{equation}

  The nonholonomic Hamilton--Jacobi equation~\eqref{eq:NHHJ} is
  \begin{equation}
    \label{eq:NHHJ-VRD}
    H \circ \gamma = E,
  \end{equation}
  where $E$ is a constant (the total energy).
  Let us construct an ansatz for Eq.~\eqref{eq:NHHJ-VRD}.
  The momentum constraint $p \in \mathcal{M}$ gives $p_{x} = m R \cos\varphi\,p_{\psi}/I$ and $p_{y} = m R \sin\varphi\,p_{\psi}/I$, and so we can write $\gamma: Q \to \mathcal{M}$ as
  \begin{multline}
    \gamma = \frac{m R}{I}\cos\varphi\,\gamma_{\psi}(x,y,\varphi,\psi)\,dx + \frac{m R}{I}\sin\varphi\,\gamma_{\psi}(x,y,\varphi,\psi)\,dy
    \\
    + \gamma_{\varphi}(x,y,\varphi,\psi)\,d\varphi+ \gamma_{\psi}(x,y,\varphi,\psi)\,d\psi
  \end{multline}
  Now we assume the following ansatz:
  \begin{equation}
    \label{eq:ansatz-VRD}
    \gamma_{\varphi}(x,y,\varphi,\psi) = \gamma_{\varphi}(\varphi).
  \end{equation}
  Then the condition $d\gamma|_{\mathcal{D}\times\mathcal{D}} = 0$ in Eq.~\eqref{eq:dgamma} gives
  \begin{equation}
    \label{eq:dgamma-VRD}
    \pd{\gamma_{\psi}}{\varphi} = 0,
  \end{equation}
  and so
  \begin{equation}
    \label{eq:ansatz1-VRD}
    \gamma_{\psi}(x,y,\varphi,\psi) = \gamma_{\psi}(x,y,\psi).
  \end{equation}
  So Eq.~\eqref{eq:NHHJ-VRD} becomes
  \begin{equation}
    \label{eq:NHHJ-VRD-1}
    \frac{1}{2}\parentheses{ \frac{\gamma_{\varphi}(\varphi)^{2}}{J} + \frac{I+m R^{2}}{I^{2}}\gamma_{\psi}(x,y,\psi)^{2} } = E.
  \end{equation}
  The first term in the parentheses depends only on $\varphi$, whereas the second depends on $x$, $y$, and $\psi$.
  This implies that both of them must be constant:
  \begin{equation}
    \gamma_{\varphi}(\varphi) = \gamma_{\varphi}^{0},
    \qquad
    \gamma_{\psi}(x,y,\psi) = \gamma_{\psi}^{0},
  \end{equation}
  where $\gamma_{\varphi}^{0}$ and $\gamma_{\psi}^{0}$ are the constants determined by the initial condition such that
  \begin{equation*}
    \frac{1}{2}\parentheses{ \frac{1}{J}(\gamma_{\varphi}^{0})^{2} + \frac{I+m R^{2}}{I^{2}}(\gamma_{\psi}^{0})^{2} } = E.
  \end{equation*}
  Then Eq.~\eqref{eq:NHHJ-curve} becomes
  \begin{equation}
    \dot{x} = \frac{\gamma_{\psi}^{0} R}{I} \cos\varphi,
    \qquad
    \dot{y} = \frac{\gamma_{\psi}^{0} R}{I} \sin\varphi,
    \qquad
    \dot{\varphi} = \frac{\gamma_{\varphi}^{0}}{J},
    \qquad
    \dot{\psi} = \frac{\gamma_{\psi}^{0}}{I},
  \end{equation}
  which are integrated easily to give the solution
  \begin{equation}
    \begin{array}{c}
      \DS x(t) = c_{1} + \frac{J R\,\gamma_{\psi}^{0}}{I\,\gamma_{\varphi}^{0}}\,\sin\parentheses{ \frac{\gamma_{\varphi}^{0}}{J}\,t + \varphi_{0} },
      \medskip\\
      \DS y(t) = c_{2} - \frac{J R\,\gamma_{\psi}^{0}}{I\,\gamma_{\varphi}^{0}}\,\cos\parentheses{ \frac{\gamma_{\varphi}^{0}}{J}\,t + \varphi_{0} },
      \medskip\\
      \DS \varphi(t) = \varphi_{0} + \frac{\gamma_{\varphi}^{0}}{J}\,t,
      \qquad
      \DS \psi(t) = \psi_{0} + \frac{\gamma_{\psi}^{0}}{I}\,t,
    \end{array}
  \end{equation}
  where $c_{1}$, $c_{2}$, $\varphi_{0}$, and $\psi_{0}$ are all constants.
\end{example}

Separation of variables for unconstrained Hamilton--Jacobi equations often deals with problems with potential forces, e.g., a harmonic oscillator and the Kepler problem.
Let us show that separation of variables works also for the following simple nonholonomic system with a potential force:
\begin{example}[The Knife Edge; see, e.g., \citet{Bl2003}]
  \label{ex:KnifeEdge}
  Consider a plane slanted at an angle $\alpha$ from the horizontal and let $(x,y)$ represent the position of the point of contact of the knife edge with respect to a fixed Cartesian coordinate system on the plane~(see Fig.~\ref{fig:KnifeEdge}).
  \begin{figure}[htbp]
    \centering
    \includegraphics[width=.65\linewidth]{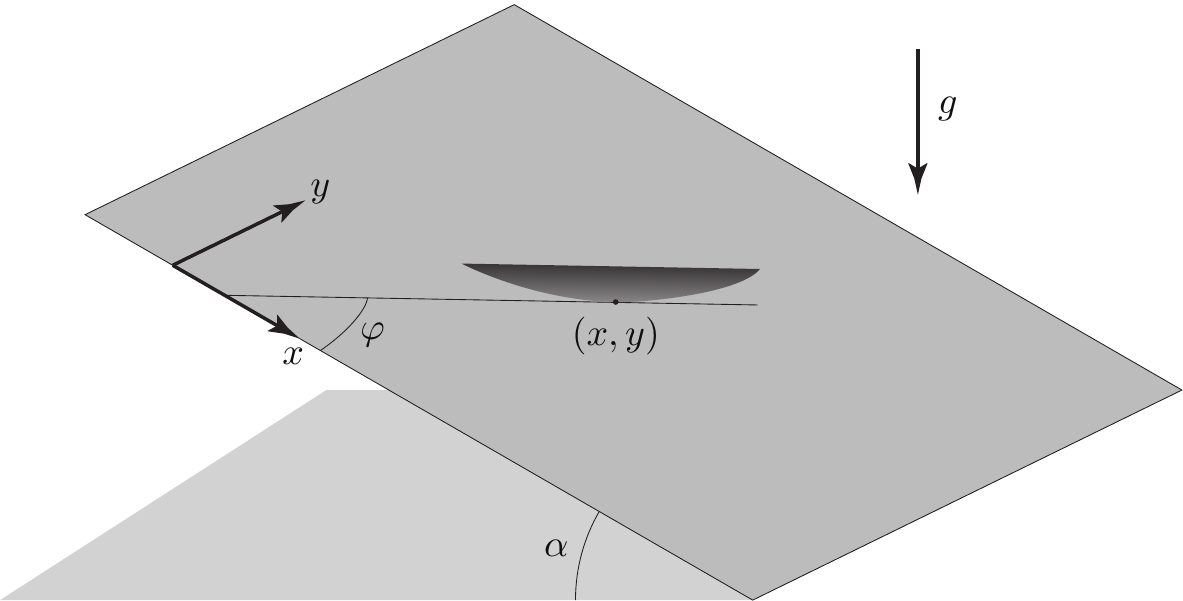}
    \caption{Knife edge on inclined plane.}
    \label{fig:KnifeEdge}
  \end{figure}
  The configuration space is $Q = SE(2) = \{ (x, y, \varphi) \}$.
  Suppose that the mass of the knife edge is $m$, and the moment of inertia about the axis perpendicular to the inclined plane through its contact point is $J$.
  The velocity constraint is
  \begin{equation}
    \label{eq:constraints-KnifeEdge}
    \sin\varphi\,\dot{x} - \cos\varphi\,\dot{y} = 0,
  \end{equation}
  and so the constraint one-form is
  \begin{equation}
    \label{eq:omega-KnifeEdge}
    \omega^{1} = \sin\varphi\,dx - \cos\varphi\,dy.
  \end{equation}
  The Hamiltonian $H: T^{*}Q \to \R$ is given by
  \begin{equation}
    H = \frac{1}{2}\parentheses{ \frac{p_{x}^{2} + p_{y}^{2}}{m} + \frac{p_{\varphi}^{2}}{J} } - m g x \sin \alpha.
  \end{equation}

  The nonholonomic Hamilton--Jacobi equation~\eqref{eq:NHHJ} is
  \begin{equation}
    \label{eq:NHHJ-KnifeEdge}
    H \circ \gamma = E,
  \end{equation}
  where $E$ is a constant (the total energy).
  Let us construct an ansatz for Eq.~\eqref{eq:NHHJ-KnifeEdge}.
  The momentum constraint $p \in \mathcal{M}$ gives
  \begin{equation*}
    p_{y} = \tan\varphi\,p_{x},
  \end{equation*}
  and so we can write $\gamma: Q \to \mathcal{M}$ as
  \begin{equation}
    \gamma = \gamma_{x}(x,y,\varphi)\,dx +  \tan\varphi\,\gamma_{x}(x,y,\varphi)\,dy
    + \gamma_{\varphi}(x,y,\varphi)\,d\varphi.
  \end{equation}
  Now we assume the following ansatz:
  \begin{equation}
    \label{eq:ansatz-KnifeEdge}
    \gamma_{\varphi}(x,y,\varphi) = \gamma_{\varphi}(\varphi).
  \end{equation}
  Then the condition $d\gamma|_{\mathcal{D}\times\mathcal{D}} = 0$ in Eq.~\eqref{eq:dgamma} gives
  \begin{equation}
    \label{eq:dgammaEq-KnifeEdge}
    \pd{\gamma_{x}}{\varphi} = -\tan\varphi\,\gamma_{x}.
  \end{equation}
  Integration of this equation yields
  \begin{equation}
    \label{eq:ansatz-KnifeEdge}
    \gamma_{x}(x,y,\varphi) = f(x,y) \cos\varphi,
  \end{equation}
  with some function $f(x,y)$.
  Then Eq.~\eqref{eq:NHHJ-KnifeEdge} becomes
    \begin{equation}
    \label{eq:NHHJ-KnifeEdge-1}
    \frac{1}{2}\brackets{ \frac{f(x,y)^{2}}{m} - (2 m g \sin\alpha)\,x + \frac{\gamma_{\varphi}(\varphi)^{2}}{J} } = E.
  \end{equation}
  The first two terms in the brackets depend only on $x$ and $y$, whereas the third depends only on $\varphi$.
  This implies that
  \begin{equation}
    \gamma_{\varphi}(\varphi) = \gamma_{\varphi}^{0}
  \end{equation}
  with some constant $\gamma_{\varphi}^{0}$, and $f(x,y)$ satisfies
  \begin{equation}
    \frac{1}{2}\brackets{ \frac{f(x,y)^{2}}{m} - (2 m g \sin\alpha)\,x + \frac{(\gamma_{\varphi}^{0})^{2}}{J} } = E.
  \end{equation}
  Let us suppose that sleigh is sliding downward in Fig.~\ref{fig:KnifeEdge}.
  Then we should have $\gamma_{x} \ge 0$ for $0 < \varphi < \pi/2$.
  From Eq.~\eqref{eq:ansatz-KnifeEdge} we see that $f(x,y) \ge 0$, and hence choose the branch
  \begin{equation}
    f(x,y) = \sqrt{ m \parentheses{ 2E - \frac{(\gamma_{\varphi}^{0})^{2}}{J} } + (2 m^{2} g \sin\alpha)\,x }.
  \end{equation}
  Then Eq.~\eqref{eq:NHHJ-curve} becomes
  \begin{equation}
    \begin{array}{c}
    \DS \dot{x} = \frac{\cos\varphi}{\sqrt{m/2}} \sqrt{ \parentheses{ E - \frac{(\gamma_{\varphi}^{0})^{2}}{2J} } + (m g \sin\alpha)\,x },
    \bigskip\\
    \DS \dot{y} = \frac{\sin\varphi}{\sqrt{m/2}} \sqrt{ \parentheses{ E - \frac{(\gamma_{\varphi}^{0})^{2}}{2J} } + (m g \sin\alpha)\,x },
    \qquad
    \DS \dot{\varphi} = \frac{\gamma_{\varphi}^{0}}{J},
  \end{array}
  \end{equation}
  Let us choose the initial condition
  \begin{equation*}
    (x(0), y(0), \varphi(0), \dot{x}(0), \dot{y}(0), \dot{\varphi}(0)) = (0, 0, 0, 0, 0, \omega),
  \end{equation*}
  where $\omega \defeq \gamma_{\varphi}^{0}/J$.
  Then we obtain
  \begin{equation}
    x(t) = \frac{g \sin\alpha}{2 \omega^{2}} \sin^{2}(\omega t),
    \qquad
    y(t) = \frac{g \sin\alpha}{2 \omega^{2}} \parentheses{ \omega t - \frac{1}{2}\sin(2\omega t) },
    \qquad
    \varphi(t) = \omega t.
  \end{equation}
  These are the solution obtained in \citet[][Section~1.6]{Bl2003}.
\end{example}

\subsection{Examples without Separation of Variables}
\label{ssec:WOSOV}
In the unconstrained theory, separation of variables seems to be the only practical way of solving the Hamilton--Jacobi equation.
However notice that separation of variables implies the existence of conserved quantities (or at least one) independent of the Hamiltonian, which often turn out to be the momentum maps arising from the symmetry of the system.
This means that the integrability argument based on separation of variables is possible only if there are sufficient number of conserved quantities independent of the Hamiltonian~\citep[See, e.g.,][\S VIII.3]{La1986}.
This is consistent with the Arnold--Liouville theorem, and as a matter of fact, separation of variables can be used to identify the action-angle variables~\citep[See, e.g.,][\S 6.2]{JoSa1998}.

The above two examples show that we have a similar situation on the nonholonomic side as well.
In each of these two examples we found conserved quantities (which are not the Hamiltonian) from the Hamilton--Jacobi equation by separation of variables as in the unconstrained theory.
So again the existence of sufficient number of conserved quantities is necessary for application of separation variables.
However, this condition can be more restrictive for nonholonomic systems since, for nonholonomic systems, momentum maps are replaced by momentum equations, which in general do not give conservation laws~\cite{BlKrMaMu1996}.

An interesting question to ask is then: What can we do when separation of variables does not seem to be working?
In the unconstrained theory, there are cases where one can come up with a new set of coordinates in which one can apply separation of variables.
An example is the use of elliptic coordinates in the problem of attraction by two fixed centers~\citep[\S47.C]{Ar1991}.
The question of existence of such coordinates for nonholonomic examples is interesting to consider.
However, we would like to take a different approach based on what we already have.
Namely we illustrate how the nonholonomic Hamilton--Jacobi theorem can be used for those examples to which we cannot apply separation of variables.
The key idea is to utilize the condition $d\gamma|_{\mathcal{D}\times\mathcal{D}} = 0$, which does not exist in the unconstrained theory as shown in Remark~\ref{rem:dgamma-2}.

\begin{example}[The Snakeboard; see, e.g., \citet{BlKrMaMu1996}]
  \label{ex:Snakeboard}
  Consider the motion of the snakeboard shown in Fig.~\ref{fig:Snakeboard}.
  \begin{figure}[htbp]
    \centering
    \includegraphics[width=.65\linewidth]{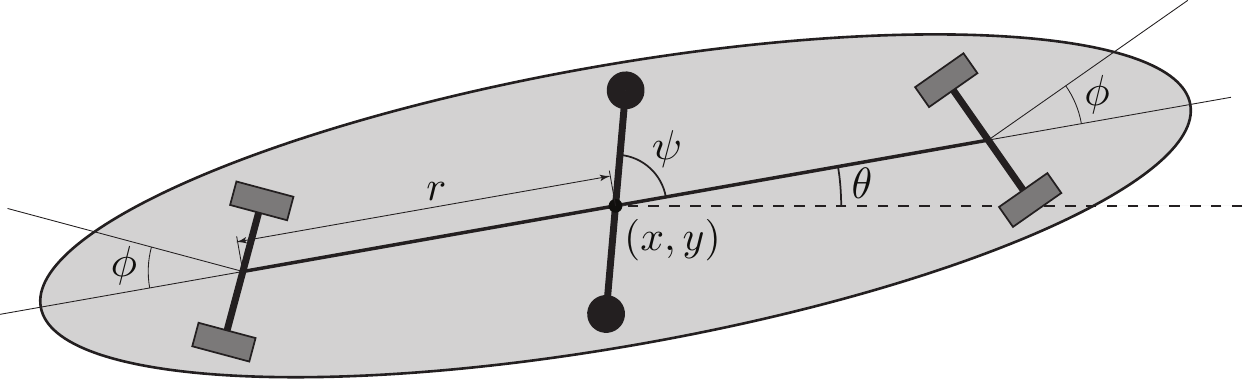}
    \caption{The Snakeboard.}
    \label{fig:Snakeboard}
  \end{figure}
  Let $m$ be the total mass of the board, $J$ the inertia of the board, $J_{0}$ the inertia of the rotor, $J_{1}$ the inertia of each of the wheels, and assume the relation $J + J_{0} + 2 J_{1} = m r^{2}$.
  The configuration space is $Q = SE(2) \times S^{1} \times S^{1} = \{ (x, y, \theta, \psi, \phi) \}$ and the Hamiltonian $H: T^{*}Q \to \R$ is given by 
  \begin{equation}
    H = \frac{1}{2m}(p_{x}^{2} + p_{y}^{2}) + \frac{1}{2J_{0}}p_{\psi}^{2} + \frac{1}{2(m r^{2} - J_{0})}(p_{\theta} - p_{\psi})^{2} + \frac{1}{4J_{1}}p_{\phi}^{2}.
  \end{equation}
  The velocity constraints are
  \begin{equation}
    \dot{x} + r \cot\phi\,\cos\theta\,\dot{\theta} = 0,
    \qquad
    \dot{y} + r \cot\phi\,\sin\theta\,\dot{\theta} = 0,
  \end{equation}
  and thus the constraint distribution is written as
  \begin{equation}
    \mathcal{D} = \setdef{ v = (\dot{x}, \dot{y}, \dot{\theta}, \dot{\psi}, \dot{\phi}) \in TQ }{ \omega^{s}(v) = 0,\, s = 1,2 },
  \end{equation}
  where
  \begin{equation}
    \omega^{1} = dx + r \cot\phi\,\cos\theta\,d\theta,
    \qquad
    \omega^{2} = dy + r \cot\phi\,\sin\theta\,d\theta.
  \end{equation}

  The nonholonomic Hamilton--Jacobi equation \eqref{eq:NHHJ} is
  \begin{equation}
    \label{eq:NHHJ-Snakeboard}
    H \circ \gamma = E.
  \end{equation}
  Let us construct an ansatz for Eq.~\eqref{eq:NHHJ-Snakeboard}.
  The momentum constraint $p \in \mathcal{M}$ gives
  \begin{equation*}
    p_{x} = -\frac{m r}{m r^{2} - J_{0}}\cot\phi\,\cos\theta\,(p_{\theta}-p_{\psi}),
    \quad
    p_{y} = -\frac{m r}{m r^{2} - J_{0}}\cot\phi\,\sin\theta\,(p_{\theta}-p_{\psi}),
  \end{equation*}
  and so we can write $\gamma: Q \to \mathcal{M}$ as
  \begin{equation}
    \gamma = 
    -\frac{m r}{m r^{2} - J_{0}}\cot\phi\,(\gamma_{\theta}-\gamma_{\psi}) (\cos\theta\,dx + \sin\theta\,dy)
    + \gamma_{\theta}\,d\theta
    + \gamma_{\psi}\,d\psi
    + \gamma_{\phi}\,d\phi
  \end{equation}
  Now we assume the following ansatz:
  \begin{equation}
    \label{eq:ansatz-Snakeboard}
    \gamma_{\psi}(x,y,\theta,\psi,\phi) = \gamma_{\psi}(\psi),
    \qquad
    \gamma_{\phi}(x,y,\theta,\psi,\phi) = \gamma_{\phi}(\phi).
  \end{equation}
  Then the nonholonomic Hamilton--Jacobi equation~\eqref{eq:NHHJ-Snakeboard} becomes
  \begin{equation}
    \label{eq:NHHJ-Snakeboard-1}
    \frac{m r^{2}}{2 (m r^{2} - J_{0})^{2}} \cot^{2}\phi\,(\gamma_{\theta} - \gamma_{\psi})^{2}
    + \frac{1}{2 J_{0}}\gamma_{\psi}^{2}
    + \frac{1}{2(m r^{2} - J_{0})}\,(\gamma_{\theta} - \gamma_{\psi})^{2}
    + \frac{1}{4 J_{1}}\gamma_{\phi}^{2}
    = E.
  \end{equation}
  Solving this for $\gamma_{\theta}$, we have
  \begin{equation}
    \label{eq:gamma_theta-Snakeboard}
    \gamma_{\theta}(x,y,\theta,\psi,\phi) =
    \gamma_{\psi}(\psi)
    +
    \frac{(m r^{2} - J_{0}) \sin\phi}{ \sqrt{ (m r^{2} - J_{0}\sin^{2}\phi)/2 } }
    \sqrt{ E - \frac{\gamma_{\psi}(\psi)^{2}}{2J_{0}} - \frac{\gamma_{\phi}(\phi)^{2}}{4J_{1}} }
  \end{equation}
 and substituting the result and Eq.~\eqref{eq:ansatz-Snakeboard} into the condition $d\gamma|_{\mathcal{D}\times\mathcal{D}} = 0$ in Eq.~\eqref{eq:dgamma} gives
  \begin{equation*}
    \begin{array}{c}
    \DS \od{}{\phi}\brackets{ \gamma_{\phi}(\phi)^{2} } = 0,
    \bigskip\\
    \DS \sin\phi \brackets{
      J_{0} \sqrt{ E - \frac{\gamma_{\psi}(\psi)}{2J_{0}} - \frac{\gamma_{\phi}(\phi)}{4J_{1}} }\, \sin\phi
      - \sqrt{ (m r^{2} - J_{0} \sin^{2}\phi)/2 }\,\gamma_{\psi}(\psi)
    } \gamma_{\psi}'(\psi) = 0.
  \end{array}
  \end{equation*}
  Therefore it follows that
  \begin{equation*}
    \gamma_{\phi}(\phi) = \gamma_{\phi}^{0},
    \qquad
    \gamma_{\psi}(\psi) = \gamma_{\psi}^{0}
  \end{equation*}
  for some constants $\gamma_{\phi}^{0}$ and $\gamma_{\psi}^{0}$.
  Hence Eq.~\eqref{eq:gamma_theta-Snakeboard} becomes
  \begin{equation*}
    \label{eq:gamma_theta-Snakeboard}
    \gamma_{\theta}(x,y,\theta,\psi,\phi) =
    \gamma_{\psi}(\psi)
    +
    \frac{(m r^{2} - J_{0})\,C \sin\phi}{ g(\phi) },
  \end{equation*}
  where we defined
  \begin{equation*}
    C \defeq \sqrt{ E - \frac{(\gamma_{\psi}^{0})^{2}}{2J_{0}} - \frac{(\gamma_{\phi}^{0})^{2}}{4J_{1}} },
    \qquad
    g(\phi) \defeq \sqrt{(m r^{2} - J_{0}\sin^{2}\phi )/2}
  \end{equation*}
  Then Eq.~\eqref{eq:NHHJ-curve} gives
  \begin{equation}
    \begin{array}{c}
      \DS \dot{x} = -\frac{C\,r \cos\theta\,\cos\phi}{g(\phi)},
      \qquad
      \DS \dot{y} = -\frac{C\,r \sin\theta\,\cos\phi}{g(\phi)},
      \medskip\\
      \DS \dot{\theta} = \frac{C \sin\phi}{g(\phi)},
      \qquad
      \DS \dot{\psi} = \frac{\gamma_{\psi}^{0}}{J_{0}} - \frac{C \sin\phi}{g(\phi)},
      \qquad
      \DS \dot{\phi} = \frac{\gamma_{\phi}^{0}}{2J_{1}}.
    \end{array}
  \end{equation}
  This result is consistent with that of \citet{KoMa1997c} obtained by reduction of Hamilton's equations for nonholonomic systems.
  It is also clear from the above expressions that the solution is obtained by a quadrature.
\end{example}

In the above example we found conserved quantities through $d\gamma|_{\mathcal{D} \times \mathcal{D}} = 0$ instead of separation of variables.
In the following example, we cannot identify conserved quantities even through $d\gamma|_{\mathcal{D} \times \mathcal{D}} = 0$; nevertheless we can still integrate the equations of motion.

\begin{example}[The Chaplygin sleigh; see, e.g., \citet{Bl2003}]
  \label{ex:ChaplyginSleigh}
  Consider the motion of the Chaplygin sleigh shown in Fig.~\ref{fig:ChaplyginSleigh}.
  \begin{figure}[htbp]
    \centering
    \includegraphics[width=.65\linewidth]{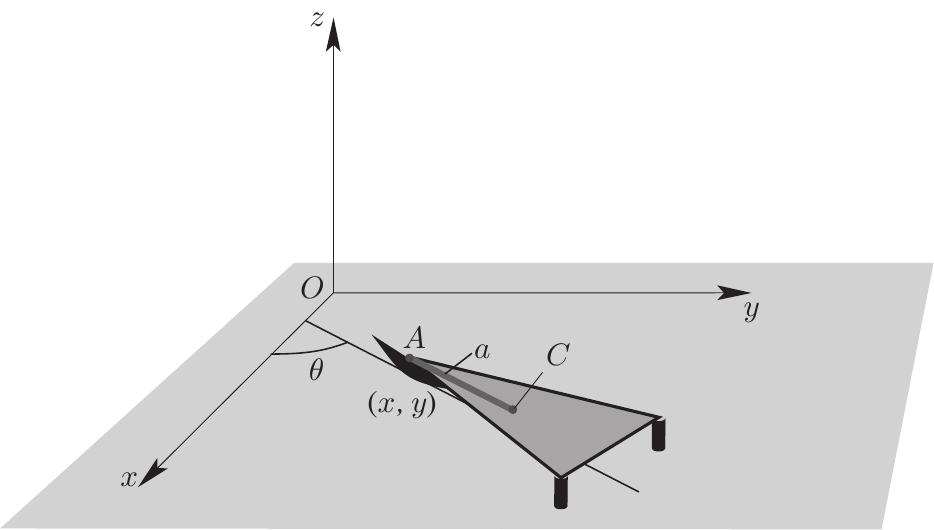}
    \caption{The Chaplygin sleigh.}
    \label{fig:ChaplyginSleigh}
  \end{figure}
  Let $m$ be the mass, $I$ the moment of inertia about the center of mass $C$, $a$ be the distance from the center of mass $C$ to the contact point $A$ of the edge.
  The configuration space is $Q = SE(2) = \{ (x, y, \theta) \}$, where the coordinates $(x,y)$ give the position of the contact point of the edge (not the center of mass).
  The velocity constraint is
  \begin{equation}
    \label{eq:constraints-ChaplyginSleigh}
    \sin\theta\,\dot{x} - \cos\theta\,\dot{y} = 0,
  \end{equation}
  and so the constraint one-form is
  \begin{equation}
    \label{eq:omega-ChaplyginSleigh}
    \omega^{1} = \sin\theta\,dx - \cos\theta\,dy.
  \end{equation}
  The Hamiltonian $H: T^{*}Q \to \R$ is given by
  \begin{multline}
    H = \frac{Ma^{2}\sin^{2}\theta + J}{2 J M}\, p_{x}^2
    +\frac{M a^{2}\cos^{2}\theta + J}{2 J M}\, p_{y}^2
    \\
    +\frac{1}{2J}\,p_{\theta}^2
    - \frac{a^{2}\sin\theta \cos\theta}{J}\, p_{x}\, p_{y}
    +\frac{a}{J}(\sin\theta\, p_{x} - \cos\theta\, p_{y})\, p_{\theta}.
  \end{multline}

  The nonholonomic Hamilton--Jacobi equation \eqref{eq:NHHJ} is
  \begin{equation}
    \label{eq:NHHJ-ChaplyginSleigh}
    H \circ \gamma = E,
  \end{equation}
  where $E$ is a constant (the total energy).
  Let us construct an ansatz for Eq.~\eqref{eq:NHHJ-ChaplyginSleigh}.
  The momentum constraint $p \in \mathcal{M}$ gives
  \begin{equation*}
  p_{y} = \tan\theta\,p_{x} + \frac{a M \sec\theta}{J + a^{2}M}\,p_{\theta},
  \end{equation*}
  and so we can write $\gamma: Q \to \mathcal{M}$ as
  \begin{equation}
    \gamma = \gamma_{x}(x,y,\theta)\,dx + \brackets{ \tan\theta\,\gamma_{x}(x,y,\theta) + \frac{a M \sec\theta}{J + a^{2}M}\,\gamma_{\theta}(x,y,\theta) } dy
    + \gamma_{\theta}(x,y,\theta)\,d\theta.
  \end{equation}
  Now we assume the following ansatz:
  \begin{equation}
    \label{eq:ansatz-ChaplyginSleigh}
    \gamma_{\theta}(x,y,\theta) = \gamma_{\theta}(\theta).
  \end{equation}
  Then the condition $d\gamma|_{\mathcal{D}\times\mathcal{D}} = 0$ in Eq.~\eqref{eq:dgamma} gives
  \begin{equation}
    \label{eq:dgammaEq-ChaplyginSleigh}
    (J + a^{2}M) \sec\theta \parentheses{ \pd{\gamma_{x}}{\theta} + \tan\theta\,\gamma_{x} } + a M \tan\theta\,\parentheses{ \od{\gamma_{\theta}}{\theta} + \tan\theta\,\gamma_{\theta} } = 0.
  \end{equation}
  On the other hand, the Hamilton--Jacobi equation \eqref{eq:NHHJ-ChaplyginSleigh} becomes
  \begin{multline}
    \label{eq:NHHJ-ChaplyginSleigh-1}
    \frac{1}{4} \sec\theta \left[
      \frac{2\sec\theta}{M}\,\gamma_{x}(x,y,\theta)^{2}
      + \frac{4a \tan\theta}{J + a^{2}M}\,\gamma_{x}(x,y,\theta)\,\gamma_{\theta}(\theta)
      \right.
      \\
      \left.
      + \frac{(J + 2a^{2}M + J \cos2\theta) \sec\theta}{(J + a^{2}M)^{2}}\,\gamma_{\theta}(\theta)^{2}
    \right]
    = E.
  \end{multline}
  It is impossible to separate the variables as we did in the examples in Examples~\ref{ex:VRD} and \ref{ex:Snakeboard}, since we cannot isolate the terms that depend only on $\theta$.
  Instead we solve the above equation for $\gamma_{x}$ and substitute the result into Eq.~\eqref{eq:dgammaEq-ChaplyginSleigh}.
  Then we obtain
  \begin{equation*}
    \od{\gamma_{\theta}}{\theta} = -a \sqrt{ M \parentheses{ 2E - \frac{\gamma_{\theta}^{2}}{J + a^{2}M} } }.
  \end{equation*}
  Solving this ODE gives
  \begin{equation}
    \label{eq:gammatheta-ChaplyginSleigh}
    \gamma_{\theta}(\theta) = (J + a^{2}M)\,\omega \cos\parentheses{ \sqrt{\frac{a^{2}M}{J + a^{2}M}}\,\theta },
  \end{equation}
  where we assumed that $x'(0) = y'(0) = 0$, $\theta(0) = 0$, and $\theta'(0) = \omega$ and also that $|\theta(t)| < \pi/2$; note that the angular velocity $\omega$ is related to the total energy by the equation $E = (J + a^{2}M)\,\omega/2$.
  Then the equation for $\theta(t)$ in Eq.~\eqref{eq:NHHJ-curve} becomes
  \begin{equation}
    \dot{\theta} = \omega \cos\parentheses{ \sqrt{\frac{a^{2}M}{J + a^{2}M}}\,\theta },
  \end{equation}
  which, with $\theta(0) = 0$, gives
  \begin{equation}
    \theta(t) = \frac{2}{b} \arctan\brackets{ \tanh\parentheses{ \frac{b}{2}\,\omega t } },
  \end{equation}
  where we set $b \defeq \sqrt{ a^{2}M/(J + a^{2}M) }$.
  Substituting this back into Eq.~\eqref{eq:gammatheta-ChaplyginSleigh}, we obtain
  \begin{equation}
    \gamma_{\theta}(t) = (J + a^{2}M)\,\omega \sech\parentheses{ \sqrt{\frac{a^{2}M}{J + a^{2}M}}\,\omega t },
  \end{equation}
  which is the solution obtained by \citet{Bl2000}~\citep[see also][Section~8.6]{Bl2003}.
\end{example}

\section{Conclusion and Future Work}
We formulated a nonholonomic Hamilton--Jacobi theorem building on the work by \citet{IgLeMa2008} with a particular interest in the application to exactly integrating the equations of motion of nonholonomic mechanical systems.
In particular we formulated the theorem so that the technique of separation of variables applies as in the unconstrained theory.
We illustrated how this works for the vertical rolling disk and snakeboard.
Furthermore, we proposed another way of exactly integrating the equations of motion without using separation of variables.

The following topics are interesting to consider for future work:
\begin{itemize}
\item {\em Relation between measure-preservation and applicability of separations of variables.}
  The integrability conditions of nonholonomic systems formulated by \citet{Ko2002} include measure-preservation.
  As mentioned above, applicability of separation of variables implies the existence of conserved quantities other than the Hamiltonian.
  Therefore it is interesting to see how these ideas, i.e., measure-preservation, applicability of separation of variables, and existence of conserved quantities, are related to each other.
  \smallskip
\item {\em ``Right'' coordinates in nonholonomic Hamilton--Jacobi theory and relation to quasivelocities.}
  In the unconstrained Hamilton--Jacobi theory, there are examples which are solvable by separation of variables only after a certain coordinate transformation.
  As a matter of fact, \citet[p.~243]{La1986} says ``The separable nature of a problem constitutes no inherent feature of the physical properties of a mechanical system, but is entirely a matter of the {\em right system of coordinates}.''
  It is reasonable to expect the same situation in nonholonomic Hamilton--Jacobi theory.
  In fact the equations of nonholonomic mechanics take simpler forms with the quasivelocities~\cite{BlMaZe2009, CoLeMaMa2009}.
  Relating the ``right'' coordinates, if any, to the quasivelocities is an interesting question to consider.
  \smallskip
\item {\em Extension to Dirac mechanics.}
  Implicit Lagrangian/Hamiltonian systems defined with Dirac structures~\cite{Va1998, YoMa2006a, YoMa2006b} can incorporate more general constraints than nonholonomic constraints including those from degenerate Lagrangians and Hamiltonians, and give nonholonomic mechanics as a special case.
  A generalization of the Hamilton--Jacobi theory to such systems is in progress~\citep{LeOhSo2009}.
\end{itemize}

\section*{Acknowledgments}
This work was partially supported by NSF grants DMS-604307 and DMS-0907949.
We would like to thank the referees for valuable comments and suggestions, and Melvin Leok, Juan Carlos Marrero, David Mart\'in de Diego, Diana Sosa, and Dmitry Zenkov for helpful discussions.

\bibliography{OhBl2009}
\bibliographystyle{plainnat}

\end{document}